\documentclass[12pt]{revtex4}
\usepackage{rotating}
\usepackage{psfig}
\usepackage{graphicx}% Include figure files

\usepackage{amsfonts}
\usepackage{amssymb}
\usepackage{amsmath}
\usepackage{latexsym}
\usepackage{rotating}

\def\beq{\begin{equation}}
\def\eeq{\end{equation}}

\def\bq{\begin{quote}}
\def\eq{\end{quote}}

\newcommand{\non}{\nonumber}
\newcommand{\be}{\begin{equation}}
\newcommand{\ee}{\end{equation}}
\newcommand{\bea}{\begin{eqnarray}}
\newcommand{\eea}{\end{eqnarray}}
\newcommand{\ba}{\begin{array}}
\newcommand{\ea}{\end{array}}

\newcommand{\la}{\lambda}

\newcommand{\vphi}{\varphi}

\newcommand{\rar}{\rightarrow}

\begin{document}

\title{\Large Anharmonic oscillator and double-well potential:
  approximating eigenfunctions}

\author{\large Alexander Turbiner\footnotemark \footnotetext{
\uppercase{E}-mail: turbiner@nucleares.unam.mx} }

\address{Instituto de Ciencias Nucleares, UNAM, Apartado Postal 70--543,
\\04510 Mexico D.F., Mexico}

\begin{flushright}
 M\'exico ICN-UNAM 05-03\\
 June 1, 2005
\end{flushright}

%\pacs{no}

\maketitle

\begin{center}
{\bf\large Abstract}
\end{center}
\small{
\begin{quote}
 A simple uniform approximation of the logarithmic derivative of the ground
 state eigenfunction for both the quantum-mechanical anharmonic oscillator
 and the double-well potential given by $V= m^2 x^2+g x^4$ at arbitrary
 $g \geq 0$ for $m^2>0$ and $m^2<0$, respectively, is presented. It is
 shown that if this approximation is taken as unperturbed problem it
 leads to an extremely fast convergent perturbation theory.
\end{quote}}

\vskip 2cm

\begin{center}
 {\it Invited contribution to Letters in Mathematical Physics \\
  to a Special Issue in memory of Professor Felix~A.~Berezin}
\end{center}

\newpage

For the last fifty years the quantum one-dimensional anharmonic
oscillator $(m^2 \geq 0)$ as well as the double-well potential
problem $(m^2 < 0)$ described by the Hamiltonian
\begin{equation}
\label{AHO}
    {\cal H}\ =\ - \frac{d^2}{dx^2}\ +\ m^2 x^2 + g x^4\ ,
\end{equation}
permanently attracted a lot of attention. The interest to these
problems ranges from various branches of physics to chemistry and
biology. It can not be an exaggeration to say that after the
seminal papers by C.~Bender-T.T.~Wu at 1969-1973 \cite{BW} in near
thousand of physics articles the problem (\ref{AHO}) was touched
in one way or another. This seemingly simple problem revealed
extremely rich internal structure which looks intrinsic for any
non-trivial problem of quantum mechanics and even for quantum
field theory. In particular, the obvious failure of the
perturbation theory in powers of $g$ due to its asymptotic
(divergent) nature in (\ref{AHO}) pushed the development of
non-perturbative methods. In practice, the anharmonic oscillator
(\ref{AHO}) served always as a test-ground for non-perturbative
methods. Another important feature of the anharmonic oscillator
(\ref{AHO}) is related to the fact that it can be interpreted as
one-dimensional quantum field theory $g \phi^4$ with zero spacial
dimension, $(1,0)$. Hence, a study of (\ref{AHO}) can be
insightful to a realistic four-dimensional quantum field theory $g
\phi^4$. The present article is devoted to the construction of a
uniform approximation of the ground state eigenfunction of
(\ref{AHO}) in $x$-space which would continue to be valid for
different values of the coupling constant $g$ and the parameter
$m^2$. If such an approximation is constructed the energy of the
ground state as well as different average values can be calculated
with a guaranteed accuracy. We are not aware of previous attempts
to construct a uniform approximation of the eigenfunctions.

Take the Schroedinger equation for (\ref{AHO})
\begin{equation}
\label{SchAHO}
    - \frac{d^2 \Psi}{dx^2}\ +\ m^2 x^2 \Psi \ +\ g x^4\Psi\ =
    \ E(m^2,g) \Psi \quad
    ,\quad \int_{-\infty}^{+\infty} |\Psi|^2 dx < \infty \ .
\end{equation}
For $m^2 \geq 0$ the potential in (\ref{SchAHO}) is a single-well
potential, which describes the celebrated quartic anharmonic
oscillator. For $m^2<0$ the potential in (\ref{SchAHO}) is a
two-well potential. It describes another celebrated potential -
the so-called double-well potential (also known as the Higgs
potential, Lifschitz potential). It is easy to check that for the
energy in (\ref{SchAHO}) the Symanzik scaling relation holds
\[
  E(m^2, g) = g^{1/3} E\bigg(\frac{m^2}{g^{2/3}}, 1\bigg)\ ,\
  \Psi (x; m^2, g) = \Psi \bigg(x g^{1/6}; \frac{m^2}{g^{2/3}}, 1\bigg)\ .
\]
This manifests that the original problem (\ref{SchAHO}) is
essentially a single-parametric problem.

Eigenfunctions of (\ref{SchAHO}) are sharply changing functions in
$x$ being characterized by a power-like behavior at $|x| \rar 0$
and an exponentially-decaying one at $|x| \rar \infty$. In order
to make them smooth we introduce the exponential representation
for eigenfunctions
\begin{equation}
\label{psi}
 \Psi (x)\ =\ e^{-\vphi (x)}\ .
\end{equation}
Following the oscillation (Sturm) theorem the function $\vphi$
should have logarithmic singularities at real $x$ which correspond
to nodes of the wavefunction. Its regular part which appears after
subtraction of those singularities should be slow-changing
function of the power-like behavior at both small and large
distances. After substitution of (\ref{psi}) into (\ref{SchAHO})
we get a Riccati equation
\begin{equation}
\label{RicAHO}
    y' - y^2\ =\ E - \ m^2 x^2\ -\ g x^4\ ,
    \quad y\ =\ \vphi'\ =\ (\log \Psi (x))'\ ,\ y(0)=0\ .
\end{equation}
This famous equation serves as a basis for developing the WKB
approximation scheme. Due to symmetric nature of the r.h.s. of
(\ref{RicAHO}) $x \rar -x$ the function $y$ is odd. Its structure
for the $n$th excited state should be
\begin{equation}
\label{phase}
 y\ =\ -\sum_{i=1}^n \frac{1}{x-x_i} \ +\ y_{reg}(x)\ ,
\end{equation}
where $x_i$ is the position of $i$th node. The regular part
$y_{reg}(x)$ should be a non-singular at real $x$ odd function
which vanish at $x=x_i,\ i=1,2,\ldots n$. It is evident that all
non-zero $x_i$ come in pairs: once we have a node at $x=x_i$
always there exists a node at $x=-x_i$.

Now we consider the case of the ground state. According to above,
in the phase (\ref{phase}) $n=0$ and singular part of $y$ is
absent, $y=y_{reg}$. Hence, the function $y$ has no singularities
at real $x$. It is easy to find its asymptotic behavior
\begin{equation}
\label{y-infty}
  y = g^{1/2} x|x| + \frac{m^2}{2g^{1/2}} \frac{|x|}{x} +
  \frac{1}{x} - \frac{4 g E + m^4}{8 g^{3/2}} \frac{1}{x|x|} -
  \frac{m^2}{2g}\frac{1}{x^3}+
  \ldots \quad \mbox{at}\ |x| \rar \infty \ ,
\end{equation}
while
\begin{equation}
\label{y-zero}
  y\ =\ E x + \frac{E^2-m^2}{3} x^3 +
  \frac{2E(E^2-m^2)-3g}{15}x^5 + \ldots \quad \mbox{at}\ |x| \rar 0
  \ .
\end{equation}
One can demonstrate that for any excited state the regular part
$y_{reg}$ has similar formal expansions at $|x| \rar 0, \infty$.

Let us develop a certain iterative procedure (perturbation theory)
for solving the Riccati equation (\ref{RicAHO}). From the point of
finding the wave function it will be a multiplicative perturbation
theory unlike a standard additive perturbation theory. Such a
perturbation theory was developed for the first time by Price
\cite{Price} and then it was rediscovered many times. Eventually,
it was called the `Logarithmic Perturbation Theory' (for the
history remarks and discussion see \cite{Turbiner:1984} and
references therein).

As a first step we choose some square-integrable function $\Psi_0$
and calculate its logarithmic derivative
\begin{equation}
\label{y0}
 y_0 = (\log \Psi_0)' = \frac{\Psi_0'}{\Psi_0}\ .
\end{equation}
It is clear that $\Psi_0$ is the exact eigenfunction of the
Schroedinger operator with a potential
\begin{equation}
\label{V0}
  V_0 = \frac{\Psi_0''}{\Psi_0} = y_0^2 - y_0'\ ,
\end{equation}
where without a loss of generality we put their eigenvalue equals
to zero, $E_0=0$. It is nothing but a choice of the reference
point for eigenvalues. Now we can construct a perturbation theory
for Riccati equation taking $\Psi_0$ and $y_0, V_0$ as zero
approximation, which characterizes the unperturbed problem. One
can write the original potential $V = m^2 x^2 + g x^4$ as a sum,
\begin{equation}
\label{VPT}
  V = V_0 + (V-V_0)\ \equiv\ V_0 + V_1\ ,
\end{equation}
thus, taking a deviation of the original potential from the
potential of the zero approximation as a perturbation. We always
can insert a formal parameter $\la$ in front of $V_1$ and develop
a perturbation theory in powers of $\la$,
\begin{equation}
\label{PT}
  E=\sum_{k=0}^{\infty} \la^k E_k \quad ,\quad y=\sum_{k=0}^{\infty}
  \la^k y_k \ ,
\end{equation}
putting $\la=1$ afterwards. Perhaps, it is worth emphasizing that
in spite of the fact that we study iteratively the equation
(\ref{RicAHO}), in general, this perturbation series has nothing
to do with a standard WKB expansion. By substituting (\ref{PT})
into (\ref{RicAHO}) we arrive at the equations which defines
iteratively the corrections
\begin{equation}
\label{PT-equation}
  {y}_k' -2 y_0 y_k = E_k - Q_k \ ,
\end{equation}
where
\begin{eqnarray*}
Q_1 &=& V_1 \ , \\
Q_k &=& -\sum_{i=1}^{k-1} y_i \cdot y_{k-i}\ ,\quad k=2,3,\ldots
\end{eqnarray*}
It is interesting that the operator in the l.h.s. of
(\ref{PT-equation}) does not depend on $k$, while $Q_k$ in the
r.h.s. can be interpreted as a perturbation on the level $k$. The
solution of (\ref{PT-equation}) can be found explicitly and is
given by
\begin{eqnarray}
 E_k & = & \frac{\int_{-\infty}^{\infty} Q_k \Psi_0^2 \,
       dx}{\int_{-\infty}^{\infty}\Psi_0^2 \, dx}\ ,
\\[10pt]
 y_k & = & \Psi_0^{-2} \int_{-\infty}^{x} (E_k - Q_k) \Psi_0^2 \,
       dx' \ .
\end{eqnarray}
It is easy to demonstrate that
\begin{equation}
\label{convergence}
 |y_1| \leq \mbox{Const}\ ,
\end{equation}
provides a sufficient condition for this perturbation theory
(\ref{PT}) to be convergent \cite{Turbiner:1984}. Note that this
condition is very rough and can be strengthened.

The first two terms in the expansion of energy (\ref{PT}) in the
above-described perturbation theory admit an interpretation in the
framework of the variational calculus. Let us assume that our
variational trial function $\Psi_0(x)$ is normalized to 1. We can
calculate the potential $V_0$ where $\Psi_0(x)$ is the ground
state eigenfunction and even put $E_0=0$ (see a discussion above).
Pure formally, we construct the Hamiltonian $H_0=p^2 + V_0$ for
which $H_0 \Psi_0(x)=0$. The variational energy is equal to
\begin{eqnarray}
\label{Evar}
  E_{var} & = & \int \psi_0 H \psi_0 = \underbrace{\int \psi_0 H_0 \,
  \psi_0}_{=E_0} +
  \underbrace{\int \psi_0 \underbrace{(H-H_0)}_{V-V_0} \psi_0}_{=E_1} \non
  \\
          & = & E_0 + E_1 (V_1=V-V_0) \geq E_{exact} \ .
\end{eqnarray}
Of course, $\Psi_0(x)$ could depend on free parameters. In this
case both $V_0$ and $V_1$ depend on parameters as well.
Minimization of $E_{var}$ with respect to the parameters can be
performed and the variational principle guarantees that $E_{var}$
gives upper bound to the ground state energy. This simple
interpretation (\ref{Evar}) reveals a fundamental difference
between perturbation theory and variational calculus. Variational
estimates can be obtained independently on the fact that the
perturbation theory associated with $\Psi_0(x)$ is convergent or
divergent. However, it seems natural to remove this difference by
requiring a convergence of the perturbation series. In this case
by calculating the next terms $E_2,E_3,\ldots$ in (\ref{PT}) one
can estimate the accuracy of variational calculation from one side
and improve it iteratively from another side. An immediate
criteria how to choose $\Psi_0(x)$ in order to get a convergent
perturbation theory is to have the perturbation potential $V_1$ to
be subordinate with respect to the non-vanishing potential of zero
approximation $V_0$,
\begin{equation}
\label{PT-converge}
 \Big|\frac{V_1}{V_0} \Big| < 1 \qquad ,
 \qquad \mbox{for}\ |x| > R \quad .
\end{equation}
This requirement has extremely non-trivial physical implication:
in order to guarantee a convergence of perturbation theory a
domain where the wavefunction is exponentially small should be
reproduced as precise as possible. The same time a description of
a domain where the wavefunction is of the order 1 is not
important. It contradicts to a straightforward physics intuition
and underlying idea of variational calculus which, in particular,
requires a precise description of the domain where the
wavefunction is of the order 1. Needless to say that namely the
latter domain gives a dominant contribution to the integrals which
define the energy in the variational calculations.

In our approach the main object for study is the logarithmic
derivative of the wavefunction $y$ (see (\ref{RicAHO})). Since $y$
is antisymmetric, we will construct different interpolations of
$y$ between $x=0$ and $x=\infty$ (see
(\ref{y-infty}),(\ref{y-zero})). In order to fulfil the
requirement of the convergence (\ref{PT-converge}) it is enough to
take into account in these interpolations the leading term of the
asymptotics at $|x| \rar \infty$. Then the interpolation we have
built is taken as zero approximation in our perturbation theory
(\ref{PT}).

{\bf 1.} The simplest interpolation can be written as follows
\begin{equation}
\label{y01}
 y_0\ =\ a x + b \sqrt{g} x |x| \ ,
\end{equation}
where $a,b$ are parameters. These parameters can be fixed either
by taking them as variational and making a minimization of
(\ref{Evar}), $a_{min},b_{min}$, or following the idea to
reproduce exactly the leading asymptotic behavior of $y$ at
$x=\infty$ (see (\ref{y-infty})), which requires to put $b=1$.

The ground state eigenfunction which corresponds to (\ref{y01})
\begin{equation}
\label{psi01}
 \psi_{0}^{(1)} = \exp \left\{ -\frac{a x^2}{2} -
                   b \frac{\sqrt{g}}{3} |x|^3
 \right\}\ ,
\end{equation}
is the exact one for the potential
\[
 V_0 = a^2 x^2 + b^2 g x^4 - 2b \sqrt{g} |x| (1-a x^2) \quad
 ,\qquad E_0 = a\ .
\]
The perturbation potential $V_1=V-V_0$ is of the form
\[
 V_1 = (m^2-a^2) x^2 + (1-b^2) g x^4 +
 2b \sqrt{g} |x| (1-a x^2)\ ,
\]
where the first two terms would disappear if we place $a=\pm m$,
and $b=1$. In this case the first two terms in the perturbation
theory are
\begin{equation}
\label{E01}
 E^{(1)} \equiv E_0+E_1 = m + 2\sqrt{g} \, \frac{\int_0^{\infty}
x(1-mx^2) e^{-mx^2 -\frac{2\sqrt{g}}{3}x^3} dx} {\int_0^{\infty}
e^{-mx^2 -\frac{2\sqrt{g}}{3}x^3} dx}\ .
\end{equation}
This expression leads to sufficiently high relative accuracy $\leq
10^{-2}$ (comparing to the accurate numerical results) for any $g
\geq 0$ and value of $m^2$. It can be confirmed by calculation of
the second correction $E_2$. However, a slight modification of of
the interpolation (\ref{y01}) by including a term $1/x$, which is
presented in (\ref{y-infty}), in a form
\begin{equation}
\label{psi01m}
 \tilde{\psi}_{0}^{(1)}\ =\ \frac{1}{\sqrt{1+c x^2}}
   \exp \left\{ -\frac{a x^2}{2} -
    b \frac{\sqrt{g}}{3} |x|^3
 \right\}\ ,
\end{equation}
where $c$ is a variational parameter, immediately leads to a
drastic increase in accuracy (see Table 1). According to a quite
simple, straightforward analysis of the second correction $E_2$
the variational energy deviates from exact one in $\lesssim
10^{-3}$ in relative units for both anharmonic oscillator and
double-well potential for studied values of the parameters in (1)
(see Table 1). When the second correction is taken into account
the relative deviation reduces in two orders of magnitude becoming
$\lesssim 10^{-5}$. It is worth mentioning  that the quality of
the approximation is reflected in the fact that $b_{min}$ deviates
from the exact value $b=1$ in several percent.

\begin{table}
\begin{center}
\begin{tabular}{|l|c|c|c|c|c|c|}
\hline &\multicolumn{2}{|c|}{$g=2$}    &
\multicolumn{2}{|c|}{$g=1$}    &
\multicolumn{2}{|c|}{$g=2$}  \\[5pt]
&\multicolumn{2}{|c|}{$m^2=1$}  & \multicolumn{2}{|c|}{$m^2=0$}  &
\multicolumn{2}{|c|}{$m^2=-1$} \\
\cline{2-7}  &  &  &  &  &  &  \\[-5pt]
             &$ b=1$ & $ b_{min}   $
             &$ b=1$ & $ b_{min}   $
             &$ b=1$ & $ b_{min}   $ \\[5pt]
             && =1.09320 && =1.13049 && = 1.19610   \\[5pt]
\hline
\hline     &  &  &  &  &  &  \\
$E^{(1)}$   &  1.6076526  &  1.6076150  &  1.0605130
            &  1.0604314  &  1.0299142  &  1.0296682   \\
            &  &  &  &  &  &  \\
\hline
            &  &  &  &  &  &  \\
$E_{2}$     & -0.0001113  &  -0.0000735 & -0.0001513
            & -0.0000692  &  -0.0003560 & -0.0001070   \\
            &  &  &  &  &  &  \\
$E^{(2)}$   &  1.6075413  &   1.6075415 & 1.0603617
            &  1.0603622  &   1.0295581 & 1.0295612    \\
            &  &  &  &  &  &  \\
\hline \hline
  Ref.\cite{Caswell:1979}    & \multicolumn{2}{|c|}{ 1.60754130 }
  & \multicolumn{2}{|c|}
      { 1.06036209} & \multicolumn{2}{|c|}{ 1.02956085}
\\ \hline
\end{tabular}
\caption{The Hamiltonian (1): Variational energy of the ground
state and its corrections in perturbation theory
(\ref{VPT}-\ref{PT}) with $\tilde{\psi}_{0}^{(1)}$ (\ref{psi01m})
as input. The results are presented for two cases: when $b=1$ (the
dominant term in asymptotics at $|x| \rar \infty$ is reproduced
exactly, see (\ref{y-infty})) and when $b$ is taken as a
variational parameter. In the latter the value of the second
energy correction $E_2$ is reduced in two times. }
\end{center}
\end{table}

The ground state function is symmetric w.r.t. $x \rar -x$. Hence,
it has an extremum at $x=0$. For any fixed $m^2 < 0$ there exists
a value $g_{crit}$ such that for $g > g_{crit}$ this extremum is a
maximum, otherwise a minimum. It is easy to find out that the
critical point $g=g_{crit}$ corresponds to the vanishing ground
state energy, $E=0$. Using the function (\ref{psi01m}) it was
calculated the critical value $g_{crit}=0.302405$ for $m^2=-1$. It
is quite interesting from physical point of view that for a family
of double-well potentials with fixed $g$ there exists a domain $0>
m^2 > (m^2)_{crit}$ where the ground-state eigenfunction has the
maximum at the origin, which corresponds to the position of the
unstable equilibrium similar to what takes place for the
single-well case. For example, if $g=1$, the value of
$(m^2)_{crit}=-2.219597$. It implies that the particle in such a
potential with the ground state energy above the barrier, $E>0$,
somehow does not feel the existence of two minima.

{\bf 2.} The expression $E^{(1)}$ (\ref{E01}) reproduces the
harmonic oscillator energy in the limit $g \to 0$. At $g=0$ the
energy $E^{(1)}$ has a singularity as it must be. However, the
expansion of (\ref{E01}) in powers of $g$ contains besides the
integer powers in $g$ also half-integer powers which must not be
present in the formal expansion of the energy in $g$ (see
(\ref{PT}). Also the interpolation (\ref{y01}) reproduces the
leading terms only in the expansions at $|x|\rar 0, \infty$. It is
a definite drawback of the interpolation (\ref{y01}) as well as
(\ref{psi01m}) and it should be fixed. One of the simplest ways to
fix it is to take the function
\begin{equation}
\label{psi02}
 \psi_{0}^{(0,0)}\ =\ \frac{1}{\sqrt{1 + c^2 x^2}}
   \exp\left\{
   -\frac{A + a x^2/2 + b g x^4/4}{(d^2 + g x^2)^{1/2}}
       \right\}\ ,
\end{equation}
where $A,a,b,c,d$ are variational parameters. If the parameters
are chosen to be
\begin{equation}
\label{parameters}
 b=\frac{4}{3}\ ,\ a=\frac{d^2}{3} + m^2\ ,
\end{equation}
the dominant and the first two subdominant terms in the asymptotic
expansion (\ref{y-infty}) are reproduced by the function
(\ref{psi02}) exactly. In this case the convergency condition
(\ref{convergence}) is satisfied, the data for $|y_1|_{max}$ are
in Table II. These two cases we call the Case 1 (two parameters
are fixed (see (\ref{parameters})) to reproduce the growing terms
in asymptotic expansion of $y$) and Case 2 (all five parameters in
(\ref{psi02}) are variational), respectively. Table II
demonstrates that even the variational energy $E^{(1)}$ already
provides extremely high accuracy: $10^{-8} - 10^{-9}$ (Case 1) and
$10^{-10} - 10^{-11}$ (Case 2) for studied values of $m^2,g$.
These results reproduce (or exceed) the best known numerical
results in literature. The order of the third energy correction
$E_3$ is already in the three-four orders of magnitude less than
the second correction $E_2$. It indicates to the extremely fast
convergence of the series in the parameter $\la$ (see (\ref{PT})).
The energies $E^{(3)}(=E^{(1)}+E_2+E_3)$ for given values of
$m^2,g$ are the most accurate among known in literature for the
moment providing 16-17 significant digits (not all these digits
are shown in Table II).

In the Case 1 we are focused to reproduce the behavior of the
wavefunction at large distances while the behavior at small
distances is defined as a result of the variational procedure. We
can calculate the coefficient in front of $x$ in the logarithmic
derivative of (\ref{psi02}) which we denote $E_{exp}$ (see Table
II). It should be the exact energy of the ground state in the case
of the exact $y$ (see (\ref{y-zero})). It is worth emphasizing
that the deviation of $E_{exp}$ from $E^{(1)}$ (or from the exact
energy $E$) appears in the fifth significant digit (!).

\begingroup
\squeezetable
\begin{sidewaystable}
\begin{center}
\begin{tabular}{|l|c|c|c|c|c|c|}
\hline
& \multicolumn{2}{|c|}{$g=2$}
& \multicolumn{2}{|c|}{$g=1$}
& \multicolumn{2}{|c|}{$g=2$}  \\[5pt]
& \multicolumn{2}{|c|}{$m^2=1$}
& \multicolumn{2}{|c|}{$m^2=0$}
& \multicolumn{2}{|c|}{$m^2=-1$} \\
\cline{2-7} &  &  &    &  & \\[-5pt]
        $b$  & $ 4/3$  & $ 1.33813399   $
             & $ 4/3$   & $ 1.33361251   $
             & $ 4/3$   & $ 1.33190626   $ \\[5pt]
 $a=c^2/3+m^2$       & 1.79502293 &    --
                     & 0.37849775 &    --
                     & 1.10179879 &    --    \\[5pt]
  $A$           & 0.7796  &    0.
             &  0.2535  &    0.
             & -1.8484  &   -2.00747  \\[5pt]
\hline
\hline      &  &  &    &  &  &  \\
$E^{(1)}$
         &  1.607541303542 &    1.607541302751 &
            1.060362094762 &    1.060362090514 &
            1.029560850845 &    1.029560831475  \\
           &  &  &  &  &  &  \\
\hline
$|y_1|_{max}$
           &  0.0026  &   --   &
            0.0029  &    --   &
            0.0064  &    --   \\
$E_{exp}$ (\ref{y-zero})
           & 1.607362918 &
           & 1.059963236 &
           & 1.030192509 & \\
\hline $E_{2}$
         &  -0.107E-08    & -0.28E-09   &
            -0.428E-08     & -0.30E-10   &
            -0.198E-07     & -0.42E-09   \\
           &  &  &&  &  &  \\
$E^{(2)}$
         & \fbox{1.607541302469}   & \fbox{1.607541302469}
         & 1.060362090485   & \fbox{1.060362090484} &
           1.029560831059   & \fbox{1.029560831054}   \\
           &  &&  &  &  &   \\
$E_{3}$
         & -0.12E-12        & -0.2E-13
         & -0.64E-12        &  0.1E-14   &
           -0.47E-11        &  0.4E-13   \\
           &  & & &  &  & \\
$E^{(3)}$
         & \fbox{1.607541302469}    & \fbox{1.607541302469}
         & \fbox{1.060362090484}    & \fbox{1.060362090484} &
           \fbox{1.029560831054}    & \fbox{1.029560831054} \\
           &  &  & & &  &    \\
\hline \hline
   Ref.\cite{Caswell:1979}
  & \multicolumn{2}{|c|}{ 1.607541303 } & \multicolumn{2}{|c|}{ 1.06036209}
  & \multicolumn{2}{|c|}{ 1.029560848}    \\ \hline
\end{tabular}
\caption{The Hamiltonian (1): Variational energy of the ground
state and its corrections in perturbation theory
(\ref{VPT}-\ref{PT}) with $\psi_{0}^{(2)}$ (\ref{psi02}) as input
(the values of the corrections $E_{2,3}$ are rounded to very few
first digits). For given $m^2, g$ the first column corresponds to
the parameters in $\psi_{0}^{(2)}$ which are chosen to reproduce
exactly the first three terms in the asymptotic expansion
(\ref{y-infty}), the second column all parameters are taken as
variational. The energies with all accurate significant digits are
framed.}
\end{center}
\end{sidewaystable}
\endgroup

The important question concerns to the deviation of the function
(\ref{psi02}) from the exact function. We consider the Case 1 when
the growing asymptotics of $y$ at large distances is reproduced
exactly. Due to this fact the perturbation theory for any studied
$m^2, g$ is fast convergent and $y_1$ is bounded for any real $x$,
\[
 |y_1|_{max} \sim 0.01\ .
\]
Hence, the function (\ref{psi02}) provides the uniform
approximation of the exact $y$. In Fig.1 for $m^2=-1, g=2$ the
behavior of $y_0$ it is shown while Fig.2 contains the behavior of
the first correction $y_1$. It is worth mentioning that the
maximal value of $y_1$: $|y_1|_{max} \sim 0.0029$ appears at $x
\approx 3.9$ where the value of $\psi_{0}^{(0,0)}$ (\ref{psi02})
is already extremely small, $\sim 10^{-9}$ and $y_0 \sim 15.2$.
Then $y_1 \propto 1/x^2$ at $x \gg 1$. The ratio $y_1/y_0$ (see
Fig.3) is the extremely small function which tends to zero at $|x|
\rar \infty$. A similar situation appears for other values of
$m^2, g$. For the double-well potential while approaching the
semiclassical limit: $m^2 \rar -\infty$ and $g$ is kept fixed,
e.g. $g = 1$ - the accuracy provided by (\ref{psi02}) remains
unchanged: $E_2 \lesssim -1. \times 10^{-6}$ for $m^2$ ranging
from -1 to -30. In the same range the parameters of (\ref{psi02})
are varied very little, being always of the same order of
magnitude, although the value of the energy changes from $\sim 1$
to $\sim -105$. It was also calculated the parameters of the
potential (1) which the ground state energy vanishes,
\[
 E (m^2=-2.2195970861, g=1)\ \approx \ 10^{-13}\ .
\]

%%%%%%%%%%%%%% FIGURE 1
\begin{figure}
\begin{center}
  \includegraphics*[width=3.in,angle=0]{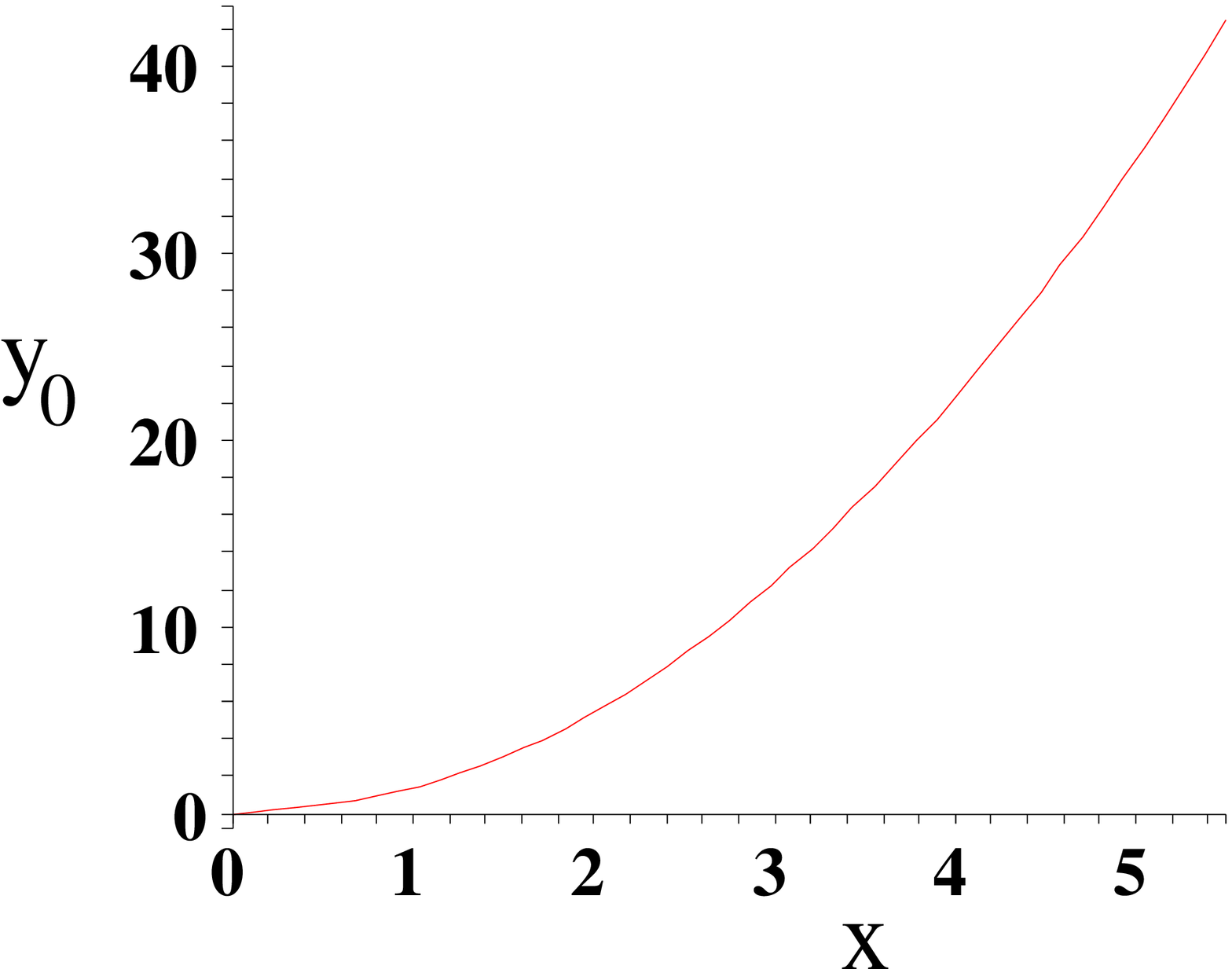}
    \caption{Logarithmic derivative $y_0$ of (\ref{psi02}) as function of
    $x$ }
  \label{fig:1}
\end{center}
\end{figure}

%%%%%%%%%%%%%% FIGURE 2
\begin{figure}
\begin{center}
  \includegraphics*[width=3.in,angle=0]{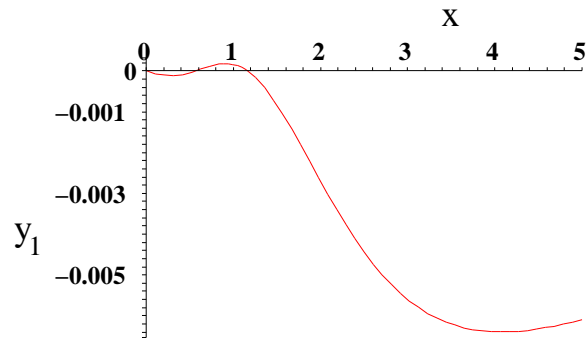}
    \caption{The first correction $y_1$ as function of
    $x$ for (\ref{psi02})}
  \label{fig:2}
\end{center}
\end{figure}

%%%%%%%%%%%%%% FIGURE 3
\begin{figure}
\begin{center}
  \includegraphics*[width=3.in,angle=0]{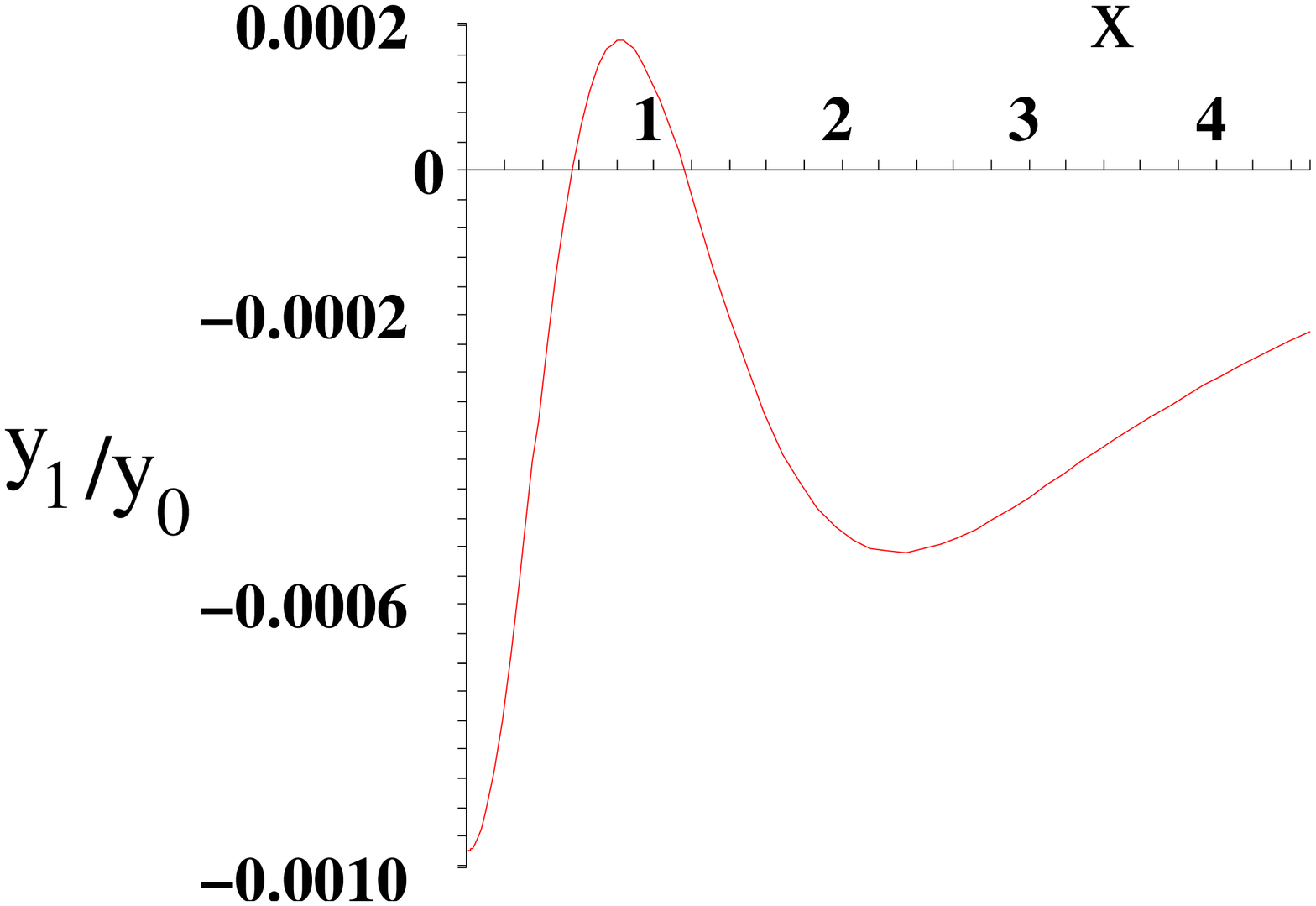}
    \caption{The ratio $y_1/y_0$ for (\ref{psi02})}
  \label{fig:3}
\end{center}
\end{figure}

It is worth mentioning that uniform approximation for
eigenfunction of $(2k+p)$th excited state, $k=0,1,2,\ldots, p=0,1$
can be easily constructed as well. It has a form
\begin{equation}
\label{psik}
 \psi_{0}^{(k,p)}\ =\ \frac{x^p P_k (x^2)}{\sqrt{1 + c^2 x^2}}
   \exp\left\{
   -\frac{A + a x^2/2 + b g x^4/4}{(d^2 + g x^2)^{1/2}}
       \right\}\ ,
\end{equation}
where $P_k$ is a polynomial of $k$th degree with positive roots.
Parameters $a,b$ are always chosen following (\ref{parameters}) in
order to reproduce the growing terms in the asymptotic expansion
of $y$ (Case 1 for the ground state)). In order to find the other
parameters the orthogonality constraint should be imposed: for
fixed $(k,p)$, (\ref{psik}) should be orthogonal to previously
constructed functions $(m,p)$, at $m=0,1,2,\ldots, k-1$. It fixes
some parameters in (\ref{psik}) while the remaining parameters are
used as variational. It is worth mentioning that very important
physical characteristic of (\ref{AHO}) at $m^2<0$ is the energy
gap between the energies of the ground state $(0,0$ and the first
excited state $(0,1$ in the semiclassical limit: $m^2$ is kept
fixed and $g \rar \infty$. Although we do not provide concrete
numerical results in this paper, this limit is reproduced with
very high accuracy.

As a conclusion I would like to recollect my first and, in fact,
the only scientific encounter with F.A.~Berezin. It was in
mid-1970s just before his breakthrough results in
supermathematics. I was a recent graduate of Moscow Institute for
Physics and Technology just hired by Institute for Theoretical and
Experimental Physics. Among theoretical physicists F.A.~Berezin
had a reputation of the unique mathematician who was able to
understand what physicists are talking about and with whom one
could discuss your mathematical difficulties. He had run a weekly
seminar at Mathematics Department of Moscow State University. By
accident I went to one of his seminars but I could not say that I
understood much. During the seminar break in the moment when I was
ready to leave he approached to me, introduced himself and asked
what I am working on. I answered that for already several years I
had been trying desperately to find an analytic solution of the
anharmonic oscillator $x^4$ (see (\ref{AHO})). Or, at least, to
understand why all my attempts had failed. He immediately said: "I
know an anharmonic oscillator, which has an analytic solution!
Take a function $e^{-x^4}$, differentiate it twice and divide the
result by $e^{-x^4}$. This is the anharmonic oscillator potential
where you know an exact eigenfunction." Then we talked for several
minutes. In the end, he said that together with M.A.~Shubin (now a
Distinguished Professor at Northeastern University in Boston) they
would write a book on the Schroedinger equation and one day if I
did not mind they could ask my opinion or advice. He introduced
M.A.~Shubin to me with whom I become a lifelong friend. After the
seminar F.A.~Berezin, M.A.~Shubin and myself took a walk to the
nearest metro station. I remember it was a very interesting
conversation about mathematics but I do not remember what it was
about - I was engrossed with thinking about the remarks of
F.A.~Berezin.

This short meeting had struck me and influenced strongly my
scientific life. From human point of view I was deeply impressed
how serious and friendly he was towards me, almost still a
student. After some time I understood that instead of solving the
Schroedinger equation with an already given potential one can
generate a zillion potentials for which a single eigenstate is
known exactly. It led me to an idea of choosing one of such
potentials as a zero approximation to construct a convergent
iterative procedure for a given eigenstate, not for the whole
spectra \cite{Turbiner:1979}. Another idea was related to a
natural question of how to construct a potential for which not one
but two or more eigenstates can be found explicitly. It led to a
discovery of quasi-exactly-solvable quantal problems
\cite{Turbiner:1988}. A few years after our meeting F.A.~Berezin
tragically died and the chance to discuss with him disappeared,
but all my life I keep a memory of him. Recently, I told this
story to one of the greatest minds in today's mathematics who was
among very close colleagues and friends of F.A.~Berezin. He
responded with sorrow that "... only after his death we realized
how strong was his influence on all of us and, eventually, how
great he was!"

%\pagebreak

{\bf Acknowledgments}

Author thanks J.C.~Lopez Vieyra the interest to the work and a
help with computer calculations. The work is supported in part by
DGAPA grant No.{\it IN124202} (Mexico).

\begingroup\raggedright
\endgroup

\end{document}